# Thermodynamic Analysis of a Polymerization-Induced Phase Separation in Nanoparticle-Monomer-Polymer Blends

**Ezequiel R. Soulé, Julio Borrajo, and Roberto J. J. Williams\***

*Institute of Materials Science and Technology (INTEMA), University of Mar del Plata and National Research Council (CONICET), J. B. Justo 4302, 7600 Mar del Plata, Argentina*

\* Corresponding author. E-mail: williams@fi.mdp.edu.ar.

ABSTRACT: We perform a thermodynamic analysis of the polymerization-induced phase separation in nanoparticle-monomer-polymer blends using a simple model recently proposed by V. V. Ginzburg (*Macromolecules* **2005**, *38*, 2362.). The model was adapted for ternary blends constituted by nanoparticles, a monomer and a linear polymer, where the relative fractions of polymer and monomer determine the conversion in the polymerization reaction. The analysis showed that phase separation can occur in the course of polymerization for a large range of values of the relevant parameters (interaction parameter, and polymer and particle sizes). This possibility has to be considered when the intention is to fix a uniform dispersion of nanoparticles in a monomer through its polymerization.

**Introduction**

One of the possible strategies to disperse nanoparticles into polymeric materials is to polymerize a homogeneous solution of these particles in the corresponding monomers. This can lead to a final homogeneous dispersion provided that phase separation does not occur in the course of polymerization. In general, when a polymerization is carried out in the presence of a second component (an oligomer, a linear polymer, a liquid crystal, etc.), phase separation can take place leading to different types of morphologies that depend on the initial composition and reaction conditions.[1] Polymerization-induced phase separation (PIPS) is used in practice to synthesize a set of useful materials such as high-impact polystyrene (HIPS),[2] rubber-modified thermosets,[3] thermoplastic-thermoset blends,[4] polymer-dispersed liquid crystals,[5] thermally-reversible light scattering films,[6,7] nanostructured thermosets,[8] etc. PIPS has also been observed in blends of epoxy monomers and polyhedral oligomeric silsesquioxanes (POSS),[9-12] that can be considered as a model system of nanoparticles dissolved in reactive solvents. A similar phenomenon has been recently reported for dispersions of inorganic nanoparticles in methyl methacrylate where aggregation of individual particles occurred upon polymerization.[13] A combination of phase separation induced simultaneously by polymerization and solvent evaporation was used to synthesize polymer-gold nanoparticle films of high dielectric constant.[14]

Our aim is to provide a thermodynamic description of the polymerization-induced phase separation in nanoparticle-monomer-polymer blends, for the particular case where a linear polymer is formed by the polymerization of the monomer. This situation is an ideal representation of a linear free-radical polymerization that produces a monodisperse polymer. The relative fractions of polymer and monomer determine the conversion in the polymerization reaction.

Among the thermodynamic models proposed to analyze nanoparticle-polymer blends, the one recently developed by Ginzburg[15] has the advantage of its simplicity and adaptability to computer simulations.[16,17] We will use this model to predict miscibility regions in ternary nanoparticle-monomer-polymer blends as a function of the relevant parameters of the system.

**Ginzburg's model**

For a binary blend consisting of nanoparticles and a linear polymer, the free energy per lattice cell ($\Delta G$) may be written as:[15]

$$\Delta G/kT = (\phi_{pol}/r_{pol}) \ln \phi_{pol} + (\phi_{part}/r_{part}) [\ln \phi_{part} + (4\phi_{part} - 3\phi_{part}^2)/(1 - \phi_{part})^2]$$
$$+ [(3R_p^2)/(2r_{pol} r_{part} R_0^2) + \chi R_0/R_p]\phi_{pol}\phi_{part} \qquad (1)$$

In eq 1, $k$ is Boltzmann constant, $T$ is temperature $\phi_{pol}$ and $\phi_{part} = 1 - \phi_{pol}$, represent the volume fractions of polymer and particles, $r_{pol}$ and $r_{part}$ are the number of lattice cells occupied by polymer and particles, $R_p$ is the radius of a spherical particle, $R_0$ is the radius of a sphere occupied by the repetitive unit of the polymer, and $\chi$ is the Flory-Huggins interaction parameter between polymer and particles.

By defining the volume of the unit lattice cell as $(4/3)\pi R_0^3$, the volume of the polymer chain with respect to the unit cell is equal to its number average degree of polymerization, $r_{pol} = N$, and the relative volume of a particle is equal to $r_{part} = (R_p/R_0)^3$. Therefore, the equilibrium conditions depend on three independent parameters: $N$, $R_p/R_0$ and $\chi$.

Changes in the arbitrary reference state used to define $\Delta G$ will incorporate constants or linear functions of compositions in eq 1. As equilibrium conditions depend on derivatives of this function, there will be no effect on thermodynamic predictions, as is obviously expected.

The first two terms of eq 1 represent the contribution of configurational entropy to free energy. Particles are assumed as hard spheres with a configurational entropy contribution described by the Carnahan-Starling equation of state.[18] The assumption of hard spheres imposes a significant penalty to the generation of a phase with a high concentration of particles (when $\phi_{part} \rightarrow 1$, $\Delta G \rightarrow \infty$).

The last term of eq 1 includes entropic and enthalpic contributions to the free energy due to interactions between polymer and particles. The entropic contribution is due to the fact that particles usually cause stretching of polymer chains in their vicinity,[15,19-21] although cases were chain contraction occurs have also been reported.[21,22] The form of the chain stretching term has its basis in the analysis of polymer brushes.[15,23] Other expressions for this term have also been suggested.[19] The enthalpic contribution depends on the interaction parameter ($\chi$) between polymer and particles. This factor includes a lumped contribution of particle-particle, particle-polymer and polymer-polymer interaction energies. It depends on temperature, e.g. it decreases with an increase in temperature for a usual upper-critical-solution-temperature (UCST) behaviour. Eventually, a dependence of $\chi$ on composition can be postulated if needed to fit the model with experimental results. Detailed analysis of nanoparticle – polymer interactions are considered in diverse models focused in the potential of mean force between a pair of spherical particles dissolved in a homopolymer melt.[24,25]

The polymer is defined as a flexible constituent that is able to occupy all the available volume even when present as a single component. In this sense, $R_0$ should be defined on the basis of the experimental density of the polymer including the free volume contribution. This is the usual way in which polymers are considered when using lattice models. However, when the size of the particle ($R_p$) gets close to the size of the repetitive unit of the polymer ($R_0$), the hard-sphere correction cannot be reconciled with the assumed flexibility of the polymer. This problem was addressed by Ginzburg,[15] who proposed an interpolating function that eliminates the hard-sphere correction in this limit. In the present analysis we avoid this limiting case and solve the equations for $R_p/R_0 \geq 3$, assuming that this value is high enough to consider the validity of the hard-sphere correction. The arbitrary interpolating function proposed by Ginzburg has an asymptotic behavior but leads to a similar situation when $R_p/R_0 \geq 3$.

Assuming typical values for the polymer mass density (~ 1 g/cm$^3$) and the molar mass of the repetitive unit (~ 100 g/mol), gives the following size of the repetitive unit: $R_0 = 0.34$ nm. This means that the present analysis is considered valid for nanoparticles with a radius equal to or higher than about 1 nm. However, there should be also a limit for the application of the model in the range of large particle sizes. Inspection of eq 1 reveals that the last term becomes negligible for large values of $R_p/R_0$. As a consequence, eq 1 predicts miscibility of large particles, a fact that is contrary to the experimental evidence. Ginzburg's model performs averaging over polymer degrees of freedom and particle degrees of freedom at once, and should not be applied for large particle sizes where a different approach should be more appropriate. Besides, increasing particle size introduces gravitational effects as a new player in the phase separation process. This driving force becomes significant for large particles when the difference between mass densities of both constituents is important. In this case, phase separation of large particles is produced by gravity independently of any thermodynamic consideration. In our analysis we restricted the range of particle sizes to $3 \leq R_p/R_0 \leq 15$.

A key issue of this analysis is whether eq 1 can be applied to describe the behaviour of a high-particle-concentration phase. According to Ginzburg,[15] eq 1 neglects the possibility of nanoparticle positional ordering or the potential effects of nanoparticle "crowding" on the polymer-particle interaction terms. In the present analysis we will assume that the nanoparticle-rich phase formed during the polymerization-induced phase separation process is amorphous. This is favoured by the fact that the Carnahan-Starling term introduces a significant entropic penalty to generate equilibrium phases containing large particle concentrations. In practice, a slight polydispersity of nanoparticle sizes should also help to avoid crystallization in the nanoparticle-rich phase. We will also assume that phases with different particle concentrations can be reversibly generated. This should be the case for several types of nanoparticles such as polyhedral oligomeric silsesquioxanes (POSS) or functionalized metal nanoparticles. The assumption could be not valid for other types of nanoparticles such as those present in fumed silica where association by chemical bonds is possible. With these considerations, the thermodynamic analysis will be performed without imposing any restriction on the concentration of particles in the system.

Before analyzing the more complex situation of a ternary blend, we will investigate the effect of these parameters on the location of miscibility regions in a binary polymer-nanoparticle blend.

**Polymer-Nanoparticle Blends**

Chemical potentials may be derived from the free energy equation in the usual form.[26] Equating chemical potentials of a particular constituent in both phases ($\alpha$ and $\beta$), leads to the following equations that define the binodal curve:

$$\ln \phi_{pol}{}^\beta/\phi_{pol}{}^\alpha + \phi_{pol}{}^\alpha - \phi_{pol}{}^\beta + (NR_0{}^3/R_p{}^3)[\phi_{part}{}^\alpha - \phi_{part}{}^\beta + 2(\phi_{part}{}^\alpha)^2(2 - \phi_{part}{}^\alpha)/(1 - \phi_{part}{}^\alpha)^2$$
$$- 2(\phi_{part}{}^\beta)^2(2 - \phi_{part}{}^\beta)/(1 - \phi_{part}{}^\beta)^2] + (R_0/R_p)(1.5 + \chi N)[(\phi_{part}{}^\beta)^2 - (\phi_{part}{}^\alpha)^2] = 0 \qquad (2)$$

$$\ln \phi_{part}{}^\beta/\phi_{part}{}^\alpha + \phi_{part}{}^\alpha - \phi_{part}{}^\beta + (R_p{}^3/NR_0{}^3)(\phi_{pol}{}^\alpha - \phi_{pol}{}^\beta) + [(8\phi_{part}{}^\beta - 5(\phi_{part}{}^\beta)^2]/(1 - \phi_{part}{}^\beta)^2$$
$$- [(8\phi_{part}{}^\alpha - 5(\phi_{part}{}^\alpha)^2]/(1 - \phi_{part}{}^\alpha)^2 + (R_p{}^2/NR_0{}^2)(1.5 + \chi N)[(\phi_{pol}{}^\beta)^2 - (\phi_{pol}{}^\alpha)^2] = 0 \qquad (3)$$

By fixing the values of $N$, $R_p/R_0$, and a particular composition in phase $\alpha$ ($\phi_{pol}{}^\alpha$ and $\phi_{part}{}^\alpha = 1 - \phi_{pol}{}^\alpha$), eqs 2 and 3 may be solved to obtain the particular composition of phase $\beta$ ($\phi_{pol}{}^\beta$ and $\phi_{part}{}^\beta = 1 - \phi_{pol}{}^\beta$) and the value of $\chi$.

The spinodal curve may be obtained following standard procedures:[26]

$$(R_p{}^3/NR_0{}^3)(1/\phi_{pol}) + (1/\phi_{part}) + 2(4 - \phi_{part})/(1 - \phi_{part})^4 - 2(R_p{}^2/NR_0{}^2)(1.5 + \chi N) = 0 \qquad (4)$$

To visualize the effect of the three parameters on the location of binodal and spinodal curves, phase diagrams in $(1/\chi)$ vs. $\phi_{part}$ coordinates were calculated (for an UCST behaviour the ordinates are proportional to temperature). Figures 1a and 1b show phase diagrams for different polymer sizes ($N$). The blend is homogeneous above the binodal curves meaning that miscibility decreases with an increase in polymer size, attaining an asymptotic value for large sizes. Equilibrium compositions are located on the binodal curves joined by horizontal tie lines. One of the branches is located at very low particle concentrations (except at compositions close to the critical point). The concentration of particles in the other branch of the binodal is limited by the entropic penalty imposed by the Carnahan-Starling (C-S) term.

Increasing the nanoparticle radius produces a significant decrease in miscibility as shown in Figure 2 for the selected range of particle sizes. The prevailing effect is the decrease in the absolute value of the contribution of configurational entropy of nanoparticles to the free energy of the system.

The effect of the C-S term may be visualized in Figure 3 where phase diagrams are predicted with and without using this term, keeping constant the rest of the parameters. The upper phase diagram is the one predicted when the C-S term is arbitrarily eliminated from the free-energy equation. The lower phase diagram is predicted using this term. The effect of the C-S term on the shape and location of the phase diagram is extremely important. In the first place it produces a significant increase in the miscibility of binary nanoparticle-polymer blends resulting from the contribution of the third term of the spinodal equation (eq 4). Besides, it shifts the critical point to the low particle concentration region leading to two equilibrium phases containing very low and intermediate particle concentrations.

The effect that is being taken into account through the C-S term may be visualized by considering that the factor $[(3R_p^2)/(2r_{pol} r_{part} R_0^2) + \chi R_0/R_p]$ in eq 1 is equal to an effective interaction parameter $\chi_{ef}$. In this case, omitting the C-S term in eq 1 would be valid for a blend of a polymer of size $N$ with a second polymer of size $(R_p/R_0)^3$, exhibiting an interaction parameter equal to $\chi_{ef}$. The blend is significantly more miscible when the second polymer is present as hard spheres instead of flexible chains. The reason is the high entropic penalty for the formation of a phase rich in hard spheres (a compact phase of hard spheres requires a significant fraction of free volume to fill the space among particles).

The higher miscibility predicted when a polymer is present as nanoparticles instead of linear chains is supported by recent experimental findings.[19] Linear polystyrene – linear polyethylene blends have an unfavourable mixing enthalpy and are a classic phase-separating system. However, branched polyethylene nanoparticles

(radius of about 13 nm) could be homogeneously dispersed in a 393-kDa linear polystyrene.[19]

**Polymerization-induced Phase Separation**

This process is modelled with a ternary blend composed of nanoparticles, a monomer and a linear polymer, where the conversion in the polymerization reaction is measured by the relative fraction of polymer in the mixture with the monomer. In fact, this is an ideal model of a linear free-radical polymerization where, at any conversion, the system is composed by high molar mass polymer and residual monomer. It is assumed that the interaction parameter between monomer and polymer is null and both components have the same interaction parameter with nanoparticles. In this situation, the free energy per lattice cell ($\Delta G$) may be written as:

$$\Delta G/kT = (\phi_{pol}/N)\ln\phi_{pol} + \phi_{mon}\ln\phi_{mon} + (\phi_{part}R_0^3/R_p^3)[\ln\phi_{part} + (4\phi_{part} - 3\phi_{part}^2)/(1 - \phi_{part})^2]$$
$$+ [(3/2N)\phi_{pol} + \chi(1 - \phi_{part})](R_0/R_p)\phi_{part} \qquad (5)$$

The equilibrium condition expressed as a binodal curve may be obtained by equating chemical potentials of the three constituents in both phases.[26] This leads to the following equations:

$$\ln\phi_{pol}^{\beta}/\phi_{pol}^{\alpha} + \phi_{pol}^{\alpha} - \phi_{pol}^{\beta} + N(\phi_{mon}^{\alpha} - \phi_{mon}^{\beta}) + (NR_0^3/R_p^3)[\phi_{part}^{\alpha} - \phi_{part}^{\beta} + 2(\phi_{part}^{\alpha})^2(2 - \phi_{part}^{\alpha})/(1 - \phi_{part}^{\alpha})^2 - 2(\phi_{part}^{\beta})^2(2 - \phi_{part}^{\beta})/(1 - \phi_{part}^{\beta})^2] + (R_0/R_p)(1.5 + \chi N)[\phi_{part}^{\beta}(1 - \phi_{pol}^{\beta}) - \phi_{part}^{\alpha}(1 - \phi_{pol}^{\alpha})] - (R_0/R_p)\chi N(\phi_{part}^{\beta}\phi_{mon}^{\beta} - \phi_{part}^{\alpha}\phi_{mon}^{\alpha}) = 0 \qquad (6)$$

$\ln \phi_{\text{mon}}{}^{\beta}/\phi_{\text{mon}}{}^{\alpha} + \phi_{\text{mon}}{}^{\alpha} - \phi_{\text{mon}}{}^{\beta} + (1/N)(\phi_{\text{pol}}{}^{\alpha} - \phi_{\text{pol}}{}^{\beta}) + (R_0{}^3/R_p{}^3)[\phi_{\text{part}}{}^{\alpha} - \phi_{\text{part}}{}^{\beta} + 2(\phi_{\text{part}}{}^{\alpha})^2(2 - \phi_{\text{part}}{}^{\alpha})/(1 - \phi_{\text{part}}{}^{\alpha})^2 - 2(\phi_{\text{part}}{}^{\beta})^2(2 - \phi_{\text{part}}{}^{\beta})/(1 - \phi_{\text{part}}{}^{\beta})^2] + (R_0/R_p)(1.5/N - \chi)(\phi_{\text{part}}{}^{\beta}\phi_{\text{pol}}{}^{\beta} - \phi_{\text{part}}{}^{\alpha}\phi_{\text{pol}}{}^{\alpha}) - (R_0/R_p)\chi[\phi_{\text{part}}{}^{\beta}(1 - \phi_{\text{mon}}{}^{\beta}) - \phi_{\text{part}}{}^{\alpha}(1 - \phi_{\text{mon}}{}^{\alpha})] = 0$ \hfill (7)

$\ln \phi_{\text{part}}{}^{\beta}/\phi_{\text{part}}{}^{\alpha} + \phi_{\text{part}}{}^{\alpha} - \phi_{\text{part}}{}^{\beta} + (R_p{}^3/NR_0{}^3)(\phi_{\text{pol}}{}^{\alpha} - \phi_{\text{pol}}{}^{\beta}) + (R_p{}^3/R_0{}^3)(\phi_{\text{mon}}{}^{\alpha} - \phi_{\text{mon}}{}^{\beta}) + [(8\phi_{\text{part}}{}^{\beta} - 5(\phi_{\text{part}}{}^{\beta})^2]/(1 - \phi_{\text{part}}{}^{\beta})^2 - [(8\phi_{\text{part}}{}^{\alpha} - 5(\phi_{\text{part}}{}^{\alpha})^2]/(1 - \phi_{\text{part}}{}^{\alpha})^2 + (R_p{}^2/NR_0{}^2)(1.5 + \chi N)[\phi_{\text{pol}}{}^{\beta}(1 - \phi_{\text{part}}{}^{\beta}) - \phi_{\text{pol}}{}^{\alpha}(1 - \phi_{\text{part}}{}^{\alpha})] + (R_p{}^2/R_0{}^2)\chi[\phi_{\text{mon}}{}^{\beta}(1 - \phi_{\text{part}}{}^{\beta}) - \phi_{\text{mon}}{}^{\alpha}(1 - \phi_{\text{part}}{}^{\alpha})] = 0$ \hfill (8)

By fixing the values of $N$, $R_p/R_0$, $\chi$, and $\phi_{\text{part}}{}^{\alpha}$, eqs 6-8 may be solved to obtain the remaining independent compositions: $\phi_{\text{pol}}{}^{\alpha}$, $\phi_{\text{pol}}{}^{\beta}$ and $\phi_{\text{part}}{}^{\beta}$ (the volume fraction of monomer in both phases is obtained by the condition that the sum of volume fractions of the three components is equal to 1).

The equation for the spinodal may be obtained following standard procedures:[26]

$(1/N\phi_{\text{pol}} + 1/\phi_{\text{mon}})[R_0{}^3/(R_p{}^3 \phi_{\text{part}}) + 1/\phi_{\text{mon}} + 2(R_0{}^3/R_p{}^3)(4 - \phi_{\text{part}})/(1 - \phi_{\text{part}})^2 - 2\chi(R_0/R_p)]$
$- [1/\phi_{\text{mon}} + 1.5(R_0/NR_p) - 2\chi(R_0/R_p)]^2 = 0$ \hfill (9)

Figure 4 shows the predicted ternary phase diagram for a monomer-polymer-nanoparticle blend with $N = 200$, $R_p/R_0 = 10$, and several values of the interaction parameter. Polymerization is simulated by horizontal trajectories starting at the particular nanoparticle concentration in the monomer-nanoparticle side and ending in the polymer-nanoparticle side. Polymerization-induced phase separation is expected for a large range of $\chi$ values located between a lower limit where the system remains homogeneous up to the end of polymerization and an upper limit where the blend is initially immiscible. For any $\chi$ value located in this range the cloud-point conversion decreases when increasing $\chi$. A large gap between binodal and spinodal curves is observed for blends with compositions in the off-critical region in the branch of high nanoparticle concentration. In this situation, a nucleation-growth type of phase separation should be favoured over spinodal demixing.

Starting from a blend containing a low nanoparticle concentration that undergoes phase separation, the final blend at the end of polymerization consists of a majority phase (the matrix) constituted practically by pure polymer, and a minority (dispersed) phase containing a higher concentration of nanoparticles than the initial blend. In practice, it is possible that polymerization-induced phase separation generates an amorphous dispersed phase rich in nanoparticles but an ordered phase is produced through a secondary phase separation inside dispersed domains when cooling from the polymerization temperature. This has been experimentally observed for blends of POSS nanoparticles in epoxy-amines monomers.[12] Amorphous POSS-rich domains were segregated at the polymerization temperature. When cooling, POSS crystals were generated inside these domains. The order-disorder transformation that occurs during cooling can be taken into account by including the free energy of the ordered phase in the thermodynamic description.

The effect of varying the polymer size (*N*) is shown in Figure 5 and the effect of particle size is shown in Figure 6. Increasing either the polymer size or the particle size produces a decrease in the conversion at which the blend enters the immiscibility region. The effect of particle size on miscibility is extremely important. A small increase in particle size from $R_p/R_0 = 3$ to $R_p/R_0 = 4$ extends the immiscibility gap to the region of very low particle concentrations. A polydispersity in the distribution of nanoparticles should lead to a fractionation by size between both phases (larger nanoparticles will be predominantly segregated to the particle-rich phase).

**Conclusions**

We discussed the possibility of producing phase separation when an initial homogeneous nanoparticle-monomer blend is polymerized with the aim of producing the dispersion of the nanoparticles in a polymeric material. The analysis was performed using Ginzburg's model and has, therefore, any limitation of this model. In the analysis we avoided the regions of very low particle sizes ($R_p/R_0 < 3$) or large particle sizes ($R_p/R_0 > 15$), where a more appropriate model is needed. The polymerization-induced phase separation was simulated using the simplest possible model in which a monomer is converted into a linear polymer with a size that does not vary with conversion (an ideal situation for a linear free-radical polymerization).

The analysis showed that phase separation can occur in the course of polymerization for a large range of values of the relevant parameters ($\chi$, $N$ and $R_p/R_0$). In most cases one of the generated phases is practically devoid of nanoparticles while the other phase exhibits a higher concentration of nanoparticles than the initial blend. This second phase constitutes the minority (dispersed) phase for the typical case of blends that contain a very low concentration of nanoparticles.

The effect of particle size on miscibility is extremely important. A small increase in particle size has a significant effect in extending the immiscibility region of the phase diagram. A polydispersity in the distribution of nanoparticles should lead to a fractionation by size between both phases (larger nanoparticles will be predominantly segregated to the particle-rich phase).

**Acknowledgment.** The financial support of the National Research Council (CONICET), the National Agency for the Promotion of Science and Technology (ANPCyT), and the University of Mar del Plata, Argentina, is gratefully acknowledged.

**Legends to the Figures**

**Figure 1.** Phase diagrams of binary polymer-nanoparticle blends for $R_p/R_0 = 5$, in ($1/\chi$) vs. $\phi_{part}$ coordinates; (a) $N = 10$ to $40$, (b) $N = 70$ to $5000$. Binodal curves are indicated by a continuous line while spinodals are traced with a dashed line.

**Figure 2.** Phase diagrams of binary polymer-nanoparticle blends for $N = 50$, in ($1/\chi$) vs. $\phi_{part}$ coordinates. Binodal curves are indicated by a continuous line while spinodals are traced with a dashed line.

**Figure 3.** Phase diagrams of binary polymer-nanoparticle blends for $N = 200$ and $R_p/R_0 = 4$, in ($1/\chi$) vs. $\phi_{part}$ coordinates. The lower phase diagram is predicted using the Carnahan-Starling term. The upper phase diagram is the prediction resulting by eliminating this term from the free-energy equation. Binodal curves are indicated by a continuous line while spinodals are traced with a dashed line.

**Figure 4.** Ternary phase diagram for a monomer-polymer-nanoparticle blend with $N = 200$, $R_p/R_0 = 10$, and several values of the interaction parameter. Binodal curves are indicated by a continuous line while spinodals are traced with a dashed line.

**Figure 5.** Ternary phase diagram for a monomer-polymer-nanoparticle blend with $\chi = 1.2$, $R_p/R_0 = 10$, and several values of the polymer length ($N$). Binodal curves are indicated by a continuous line while spinodals are traced with a dashed line.

**Figure 6.** Ternary phase diagram for a monomer-polymer-nanoparticle blend with $\chi = 1.2$ and $N = 200$. Binodal curves are indicated by a continuous line while spinodals are traced with a dashed line.

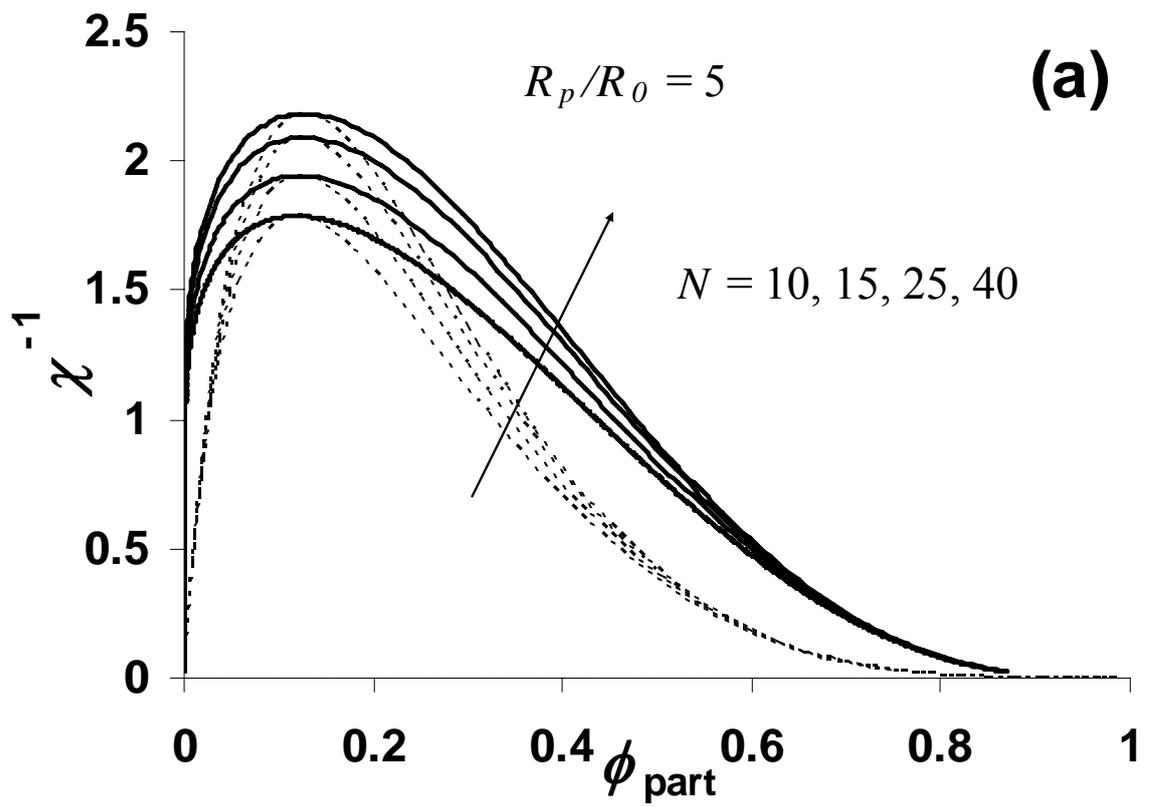

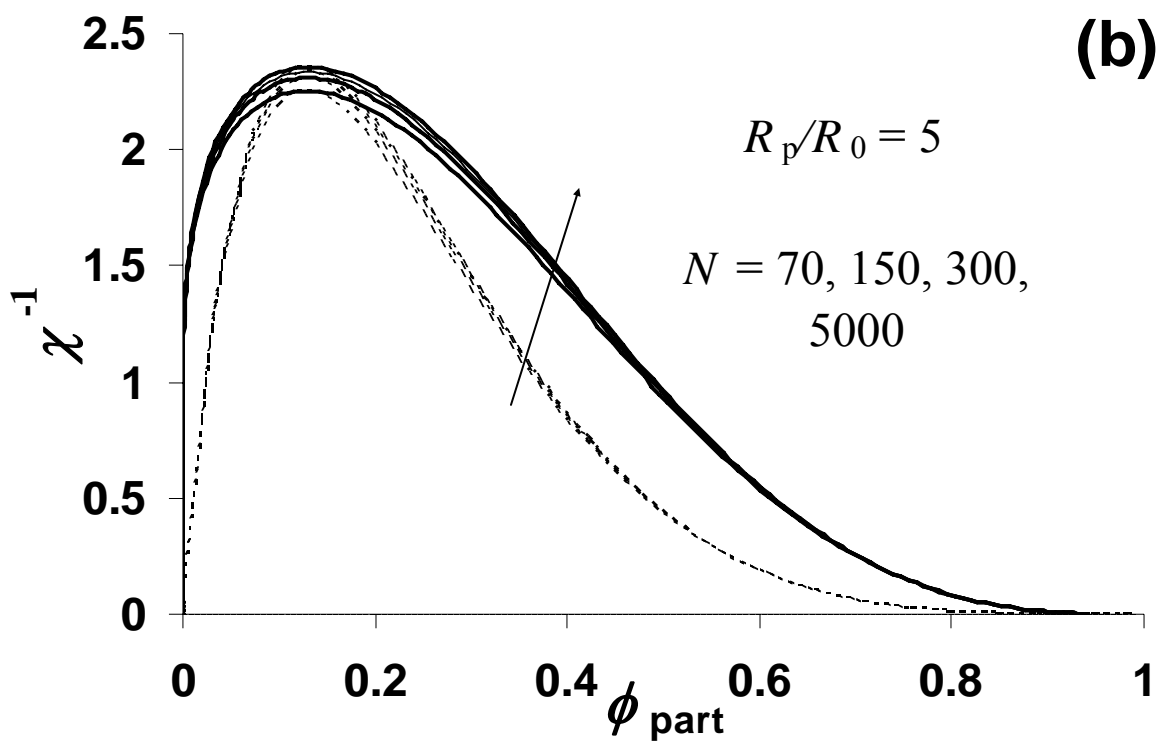

**Figure 1**

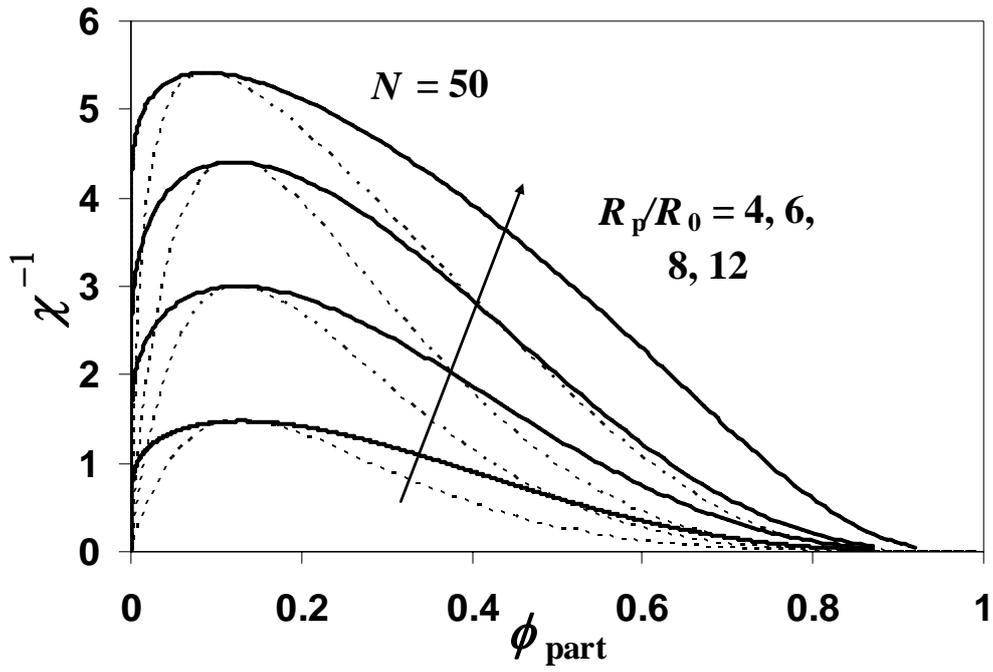

**Figure 2**

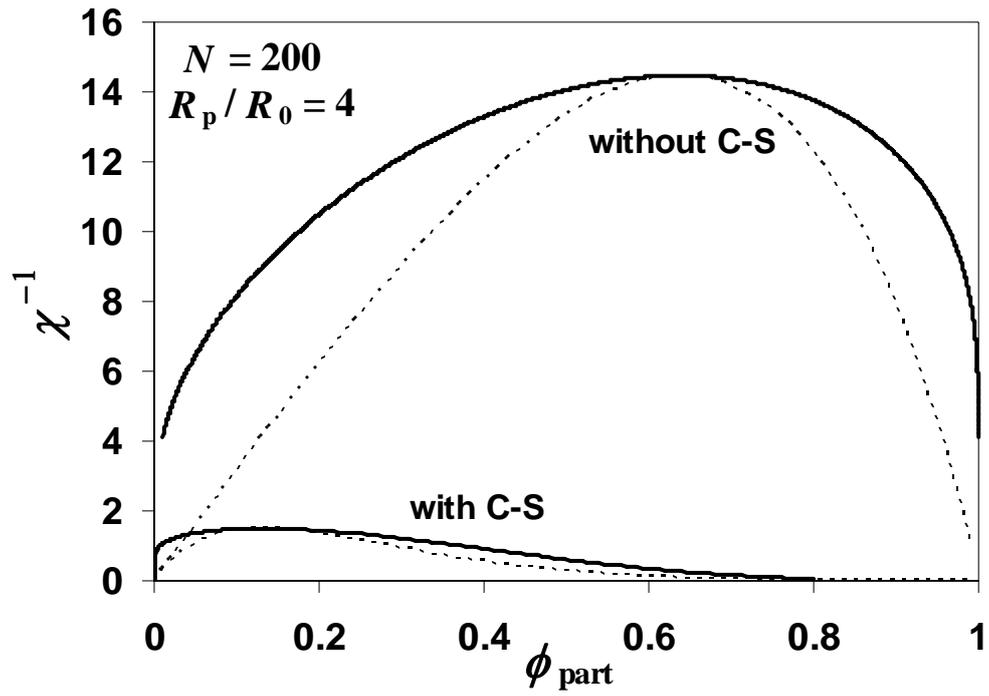

**Figure 3**

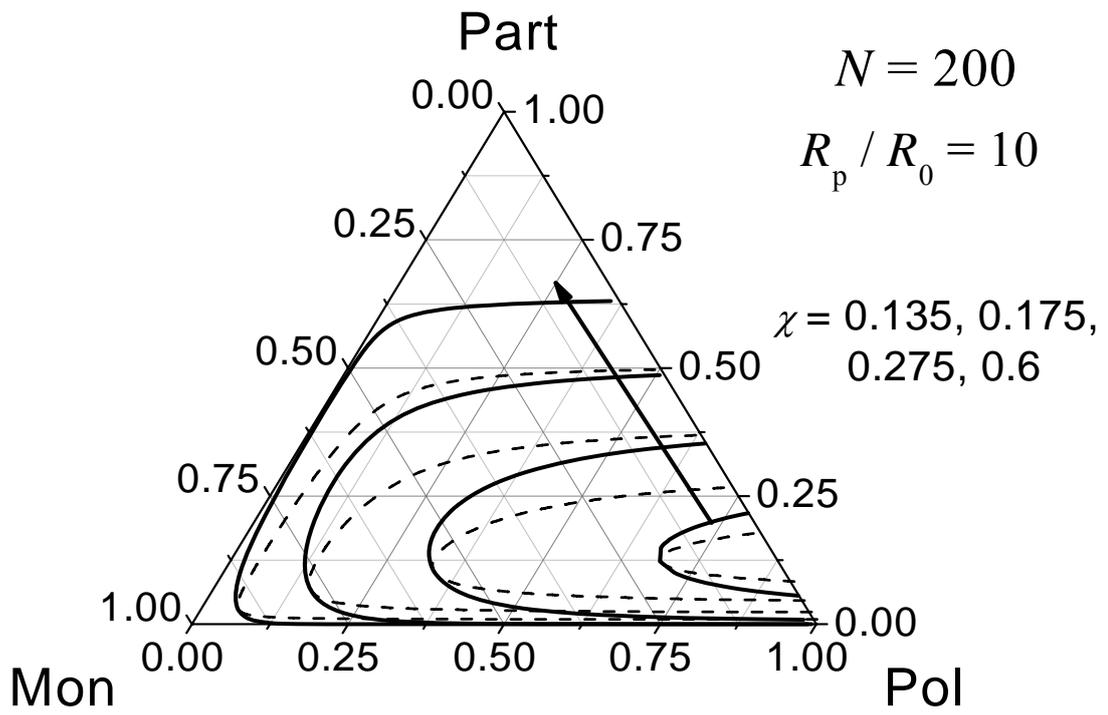

**Figure 4**

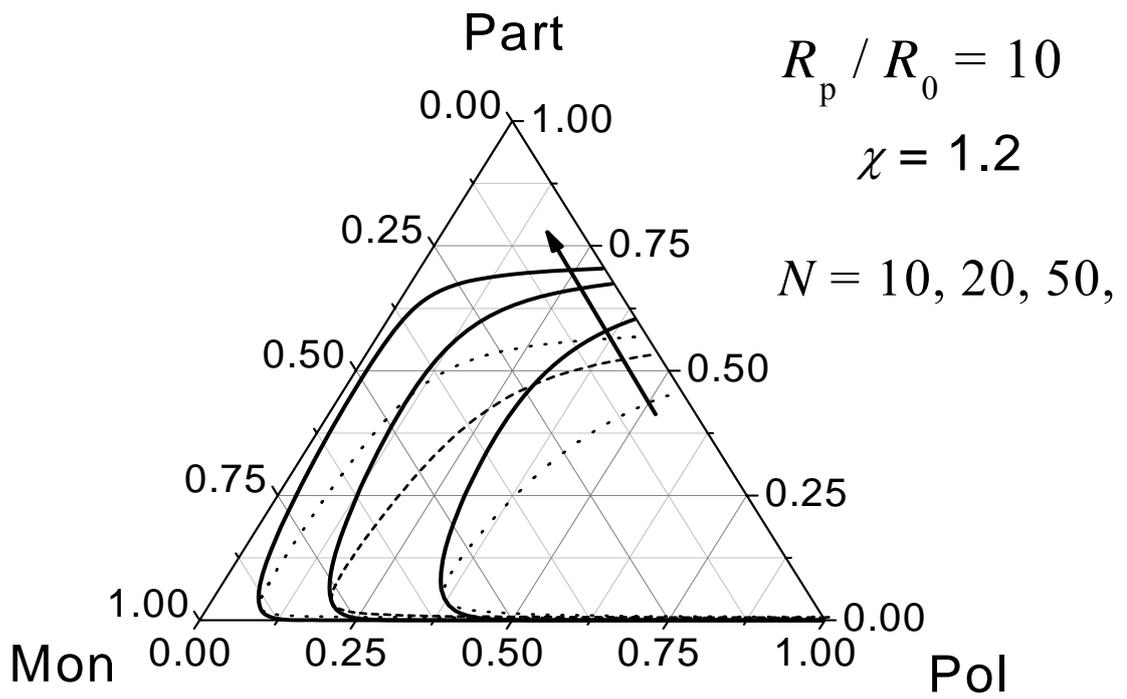

**Figure 5**

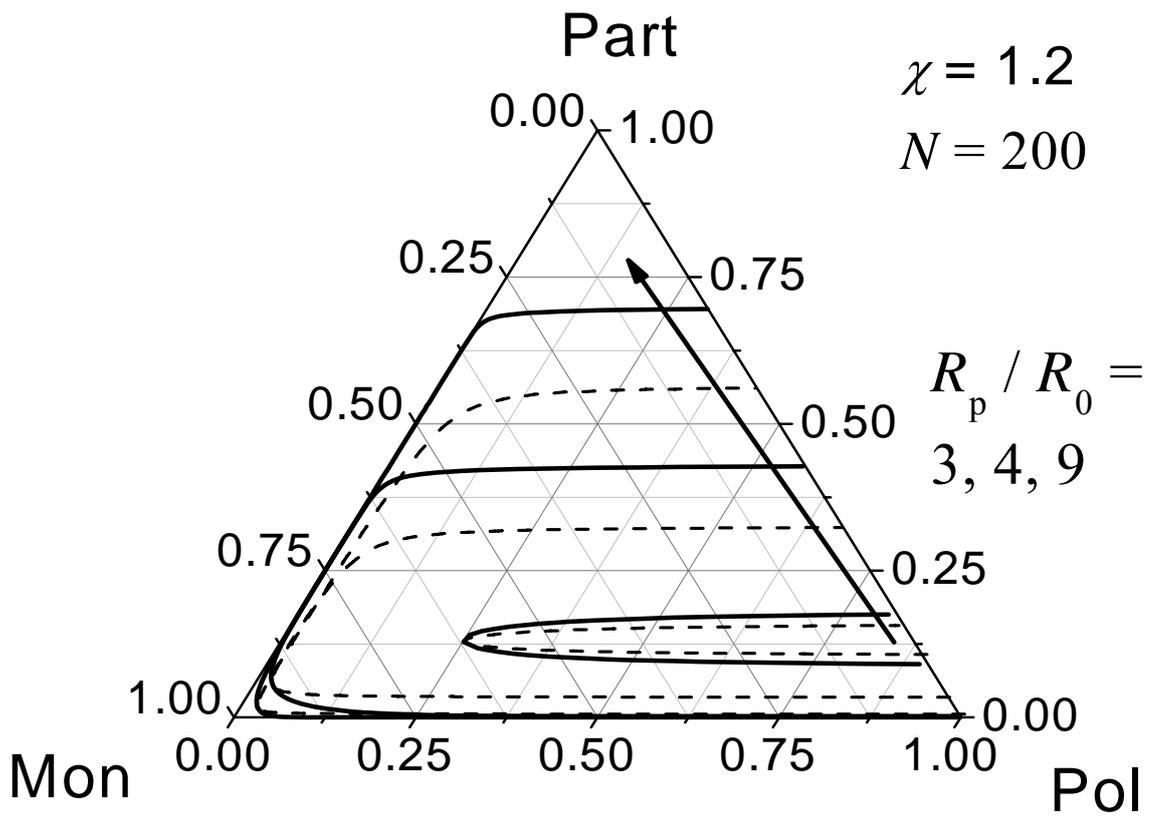

Figure 6

**For Table of Contents use only**

Thermodynamic Analysis of a Polymerization-Induced Phase Separation in Nanoparticle-Monomer-Polymer Blends

**Ezequiel R. Soulé, Julio Borrajo, and Roberto J. J. Williams***

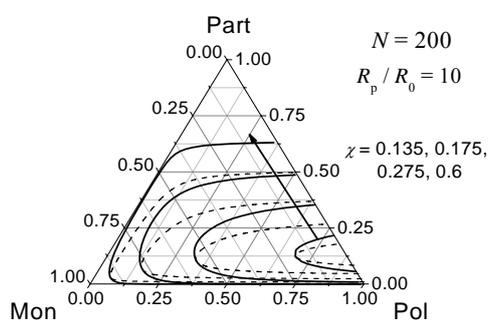